# EL-DIP: A Deep Image Prior Framework for Accurate, Efficient, and Noise-Robust Elastography

**Hao Liu**[1,2,†], **Fuyi Li**[3,†], **Dong Liu**[4], **Yue Mei**[1,2,*], **Sheng Mao**[3,*]

[1]Department of Engineering Mechanics, Dalian University of Technology, Dalian, 116023, PR China

[2]International Research Center for Computational Mechanics, Dalian University of Technology, Dalian 116023, PR China

[3]School of Mechanics and Engineering Science, Peking University, Beijing 100871, PR China

[4]School of Biomedical Engineering and Suzhou Institute for Advanced Research, University of Science and Technology of China, Suzhou, 215123, China

[†]These authors contributed equally to this work

[*]Corresponding authors: meiyue@dlut.edu.cn, maosheng@pku.edu.cn

## Abstract

Elastography is a key technique for non-invasive imaging, yet it remains severely ill-posed and sensitive to measurement noise, causing blurring and artifacts. We propose EL-DIP, a training-data-free framework based on deep image prior (DIP) that reconstructs heterogeneous shear modulus distributions. Inverse distance weighting interpolation enables arbitrary-shaped domains, and adjoint sensitivity analysis provides efficient gradient computation. Numerical and experimental studies show that EL-DIP accurately recovers inclusion geometry and modulus across different noise levels. Under extreme noise, image fidelity improves by approximately 97.1% over conventional methods. We believe that this method has great potential as a robust tool for biomedical diagnosis, biomechanical characterization, and non-destructive evaluation.

## Teaser

EL-DIP reconstructs material properties from noisy measurements without paired data.

## Introduction

Elastography is a non-invasive imaging technique for reconstructing the spatial distribution of mechanical properties, such as shear modulus and Young's modulus, inside a target domain from measured mechanical response fields, such as displacement and strain fields, under external loading[1–3]. Because pathological tissues, material defects, or distinct microstructural regions often exhibit mechanical properties that differ markedly from those of the surrounding background, elastography has become an important tool in biomedical imaging[4–8] and has also been explored for material characterization[9–11] and structural health monitoring[12,13].

In elastography, recovering the spatial distribution of elastic properties from measured deformation fields is commonly formulated as a constrained optimization problem[14–17]. However, this inverse problem is severely ill-posed and highly sensitive to measurement noise, such that small perturbations in the measured data may lead to pronounced oscillations or artifacts in the reconstructed modulus field[14,18]. To stabilize the inversion, conventional methods typically introduce explicit regularization terms, such as Tikhonov or total-variation regularization, to impose prior constraints on the solution[15,16,19,20]. Under strong noise contamination, a larger regularization parameter is often required to suppress noise-induced oscillations. However, excessive regularization may oversmooth material interfaces, thereby degrading the reconstruction of both the geometric features and elastic properties of internal inclusions[15]. As a result, the overall reconstruction quality can deteriorate significantly.

Recent advances in deep learning have introduced new computational paradigms for solving inverse problems[21,22]. In elastography, learning-based methods, including convolutional neural netwoks (CNNs)[23,24], conditional generative adversarial networks (cGANs)[25–27], and physics-informed neural networks (PINNs)[28,29], have been explored to reconstruct elastic property distributions from measured mechanical responses. Among these methods, CNN- and cGAN-based methods typically learn the mapping between measured response fields and elastic property distributions from data, whereas PINN-based methods incorporate governing equations and boundary conditions into the learning process. Although these methods have shown promising performance, they often require paired training datasets, representative training samples, or carefully designed physical constraints. Their performance may therefore degrade when the geometry, loading condition, noise level, or material distribution differs from those represented during training. This limitation is particularly restrictive in biomedical and experimental settings, where high-quality labeled datasets are often difficult to obtain.

Different from these learning-based reconstruction models, deep image prior (DIP) exploits the architecture of an untrained concolutional neural network as an implicit regulaizer[30–32]. Instead of pretraining a model on an external dataset, DIP represents the unknown image or field using a untrained neural network and optimizes the network parameters while solving exact inverse problem. The network architecture tends to recover coherent low-frequency structures before fitting high-frequency noise, which makes DIP attractive for ill-posed inverse problems with noisy observations. DIP-based reconstruction has been successfully explored in inverse imaging problems such as electrical impedance tomography (EIT)[33] and diffraction tomography[34], suggesting its potential for noise-robust elastography.

Motivated by this property, we propose a novel DIP-driven elastography method to tackle the challenges of heterogeneous elastography in high-noise environments. The explicit regularization terms in conventional methods are replaced by leveraging the implicit regularization inherent in the neural network architecture, and the network is integrated with a finite element solver to form a unified framework. The capability of the method to accurately identify the geometric shape and shear modulus of inclusions in noisy environments is demonstrated using both numerical and experimental data, along with the exhibition of favorable robustness. An overview of the framework is shown in Fig. 1.

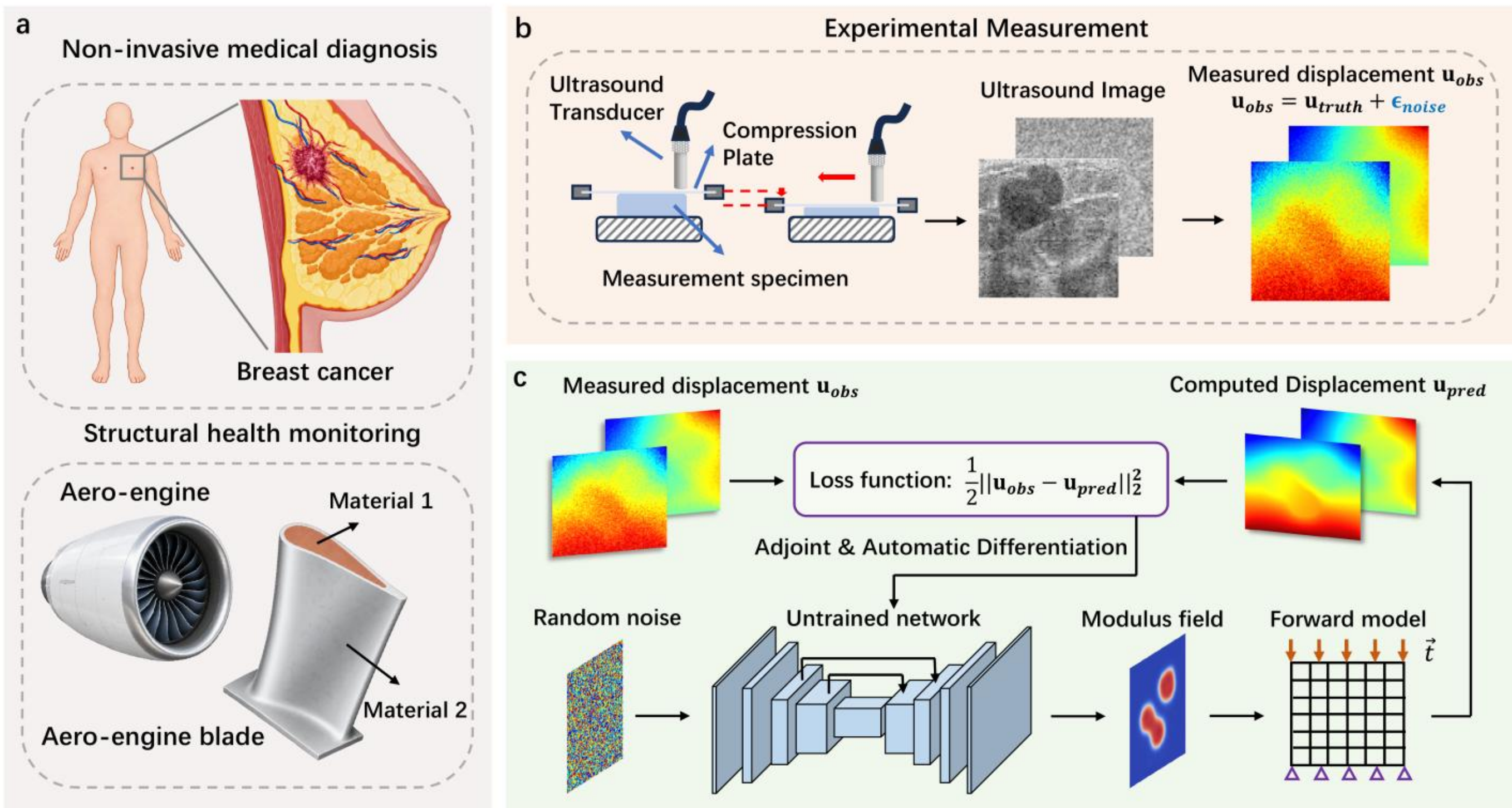

**Fig. 1. Overview of the EL-DIP framework for heterogeneous elasticity reconstruction. (a)** Representative applications in non-invasive medical diagnosis and structural health monitoring. **(b)** Experimental measurement of deformation fields under mechanical loading, illustrated using ultrasound imaging; the resulting displacement field contains inevitable measurement noise. **(c)** EL-DIP reconstructs heterogeneous elastic properties from measured displacements using an untrained neural network coupled with a physics-based forward model and adjoint-based optimization.

The rest of the paper is organized as follows. We start by introducing the elastography inverse problem and the proposed EL-DIP framework, including the finite-element forward model, DIP-based parameterization, IDW mapping, and the adjoint-based optimization strategy. The detailed adjoint derivation and implementation settings are provided in the Supplementary Materials. We then evaluate the proposed method using dimensionless numerical benchmarks and experimental data under different noise levels, with comparisons against a conventional optimization-based method. Finally, we discuss the spectral-bias mechanism underlying the noise robustness of EL-DIP, together with its limitations and future directions.

## Method

### Forward problem

The forward problem in elasticity is concerned with the calculation of the deformation within solid under external forces, assuming that the material distribution of the domain is known. The governing equations are given as follows:

$$
\begin{aligned}
\nabla \cdot \boldsymbol{\sigma} &= 0 \quad \text{in } \Omega, \\
\mathbf{u} &= \mathbf{u}_0 \quad \text{on } \Gamma_u, \\
\boldsymbol{\sigma} \cdot \mathbf{n} &= \mathbf{t}_0 \quad \text{on } \Gamma_t.
\end{aligned} \tag{1}
$$

In this work, $\Omega$ represents the entire domain of interest, $\Gamma_u$ denotes the displacement boundary, and $\Gamma_t$ refers to the traction boundary, such that $\Gamma_u \cup \Gamma_t = \partial\Omega$ and $\Gamma_u \cap \Gamma_t = \emptyset$. $\mathbf{u}_0$ and $\mathbf{t}_0$ represent the displacement and traction force on the boundary, respectively. $\mathbf{n}$ is the outward normal vector of the traction boundary, $\boldsymbol{\sigma}$ is the stress tensor, and $\mathbf{u}$ is the displacement field.

For an isotropic linear elastic material, the small-strain tensor is defined as $\boldsymbol{\epsilon} = \frac{1}{2}(\nabla\mathbf{u} + \nabla\mathbf{u}^T)$

, and in the compressible case, the constitutive relation is given by $\boldsymbol{\sigma} = 2\mu\boldsymbol{\epsilon} + \lambda\mathrm{tr}(\boldsymbol{\epsilon})\mathbf{I}$,

where $\mu$ and $\lambda$ are the Lamé parameters, and $\mathbf{I}$ is the identity tensor.

For soft tissue-like materials, the deformation is commonly assumed to be nearly incompressible. In the incompressible limit, the volumetric strain is constrained by $\mathrm{tr}(\boldsymbol{\epsilon}) = \nabla \cdot \mathbf{u} = 0$.

Therefore, a pressure-like Lagrange multiplier $p$ is introduced to enforce the incompressible constraint, and the stress tensor can be written as:

$$
\boldsymbol{\sigma} = 2\mu\boldsymbol{\epsilon} - p\mathbf{I}. \tag{2}
$$

This incompressible linear elastic model is adopted in this work. Thus the displacement can be obtained by the finite element method (FEM) leading to the following linear algebra system $\mathbf{K}\tilde{\mathbf{u}} = \tilde{\mathbf{f}}$, where $\mathbf{K}$ represents the stiffness matrix formed after discretization of problem domain, the nodal displacement vector are represented by the vector $\tilde{\mathbf{u}}$ in the discrete system, and $\tilde{\mathbf{f}}$ is the nodal force vector. Therefore, for a given shear modulus distribution, the nodal displacements are obtained by solving the corresponding system of linear equations.

### Elastography based on DIP

The inverse problem in this work is solved by using the known real nodal displacement data $\tilde{\mathbf{u}}'$ from the problem domain to map the shear modulus distribution. The solution to this inverse problem is usually obtained by minimizing the following objective function:

$$
\hat{\mu} = \underset{\mu}{\mathrm{argmin}} \left[\frac{1}{2} \parallel \tilde{\mathbf{u}}' - \tilde{\mathbf{u}}(\mu) \parallel^2 + \alpha \mathrm{Reg}(\mu)\right]. \tag{3}
$$

To avoid the overfitting issue, a regularization term $\mathrm{Reg}(\mu)$ is explicitly added as prior knowledge to constrain the elastography reconstruction process. The regularization factor $\alpha$ controls the contribution of the regularization term to the objective function. An efficient way to solve this inverse problem is based on the adjoint method and more details can be seen in[15,35].

In our method, the shear modulus distribution is reconstructed using DIP framework. The minimization problem of the EL-DIP framework can be expressed as:

$$
\begin{aligned}
J(\theta) &= \frac{1}{2} \parallel \tilde{\mathbf{u}}' - \tilde{\mathbf{u}}(f(\theta, z)) \parallel^2, \\
\hat{\theta} &= \underset{\theta}{\mathrm{argmin}} J(\theta), \\
\hat{\mu} &= f(\hat{\theta}, z).
\end{aligned} \tag{4}
$$

where $f(\theta, z)$ denotes an untrained convolutional neural network parameterized by $\theta$, and $z$ is a fixed random noise input. Unlike the classical optimization formulation for elastography, which directly updates the shear modulus field and usually requires an explicit regularization term, the DIP-based formulation optimizes the network parameters $\theta$. The reconstructed shear modulus field is then obtained from the optimized network output $f(\hat{\theta}, z)$.

### Neural network architecture

Following the network design commonly used in DIP, a U-Net-like encoder-decoder architecture is adopted as the backbone network of the EL-DIP framework[30,36]. The input of the network is random noise, sampled from the uniform distribution $z \sim U(0,0.1)$, with a size of $1 \times 96 \times 96$. The skip connections are configured as $n_s = [0,0,0,4,4]$, where the value of $n_s[i]$ denotes the number of $1 \times 1$ convolutional kernels in the i-th skip connection.

If the value is 0, there will be no skip connection. The convergence process of the model is stabilized through the fusion of feature maps from the encoder (downsampling path) and the decoder (upsampling path) via skip connections. The downsampling in the network is achieved through the stride within convolutional layers (with a kernel size of $3 \times 3$), while bilinear upsampling is employed for the upsampling process. In the network, the LeakyReLU activation function is employed in the network to enhance its nonlinear representational capacity[37]. The convolutional neural network used in this paper is shown in the Fig. 2a.

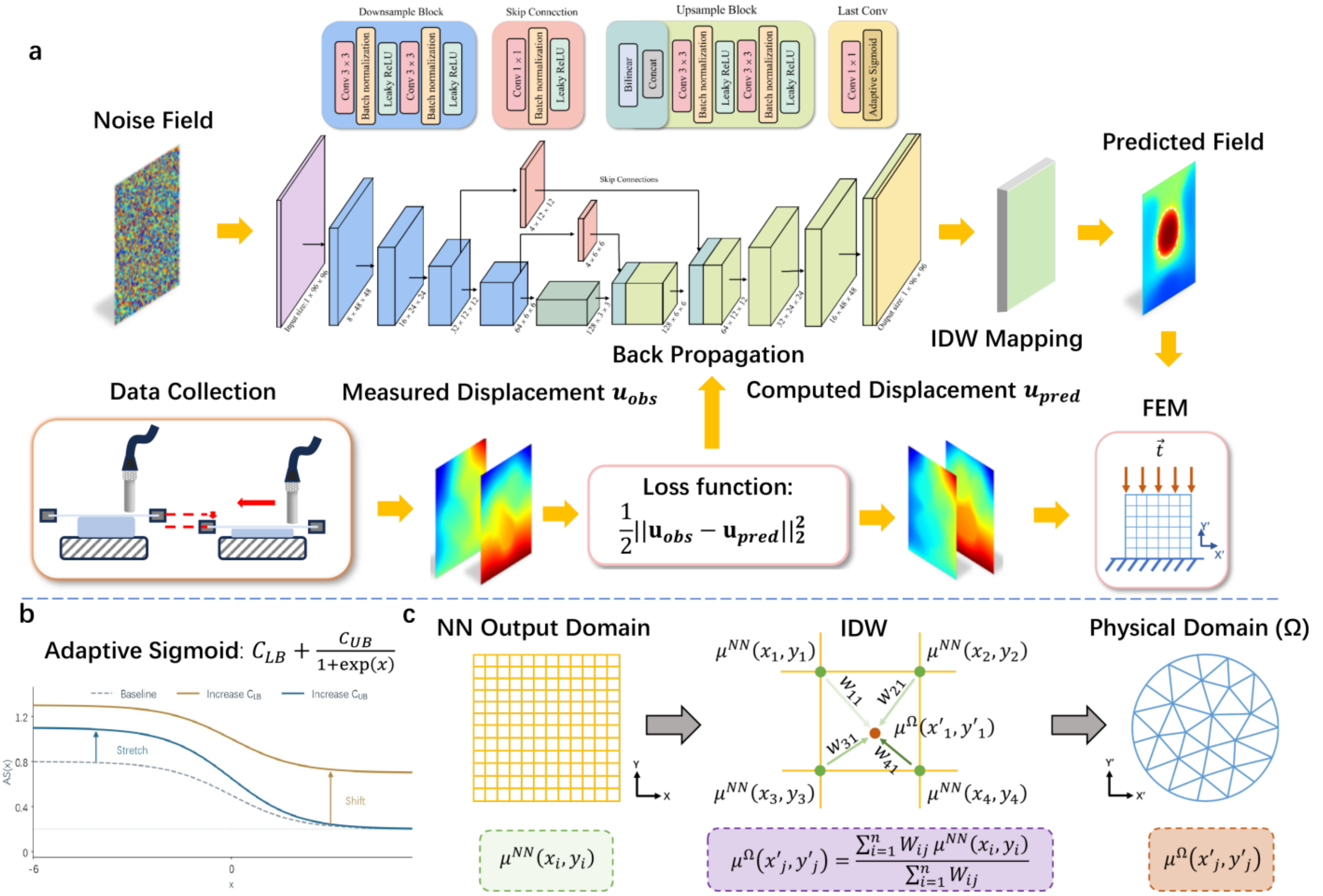

**Fig. 2. Network architecture and physics-based reconstruction. (a)** The encoder–decoder network is coupled to IDW mapping and a finite element solver; the displacement-misfit loss is back-propagated to update the network. The inset identifies the convolution, normalization, activation, and sampling blocks. **(b)** Adaptive sigmoid output scaling. **(c)** IDW interpolation from the regular network grid to nodes in rectangular or arbitrarily shaped physical domains.

In the last layer of the network output, we use the sigmoid activation function to prevent the predicted shear modulus distribution from having non-positive values. However, it is not sufficient to simply add a common sigmoid function. In fact, the shear modulus does not lie solely between (0, 1). Thus, we need to set new upper and lower limits for the sigmoid function. But in practical applications, we cannot directly know the upper limit of the shear modulus in the problem domain, so it is usually impossible to set this value manually. To address this issue, we apply an adaptive sigmoid activation layer to scale the network output $AS(x) = \frac{C_{UB}}{1+\exp(x)} + C_{LB}$, which is shown in the Fig. 2b. $C_{UB}$ and $C_{LB}$ are defined as learnable upper and lower bounds for the shear modulus, respectively, which are adaptively learned during training.

### Arbitrary shape mapping

The convolutional neural network (CNN) is inherently well-suited for feature extraction on rectangular problem domains. However, in practical scenarios, it cannot be guaranteed that the problem domain is rectangular. To enhance the generalization ability of our method, we introduce the inverse distance weighting (IDW) interpolation method[38], which maps the shear modulus $\mu^{NN}$ output from the network to the discretized nodes of the actual problem domain, as shown in Fig. 2c.

Let $\mu^{NN}(x_i, y_i)$ denote the network-predicted shear modulus value at i-th output grid point $(x_i, y_i)$. The coordinates $(x_i, y_i)$ are assigned by embedding the regular output grid into the bounding box of the problem domain, so that the network output grid and the finite element mesh are described in the same normalized coordinate system. Similarly, the shear modulus distribution of the problem domain is represented as $\mu^{\Omega}(x_j', y_j')$, where $x_j'$ and $y_j'$ are the coordinates of the j-th node on the actual grid.

For each finite element node, the mapped shear modulus is computed as a weighted average of all network-output values. The contribution of each output point is determined by its distance to the target finite element node. Specifically, for the j-th finite element node, the distance from the i-th network output point and the correponding inverse-distance weight are defined as:

$$d_{ij} = \sqrt{(x_i - x_j')^2 + (y_i - y_j')^2}, \quad w_{ij} = \frac{1}{d_{ij}^h}. \tag{5}$$

Using these weights, the shear modulus value at the j-th finite element node s obtained by the normalized weighted average:

$$\mu^{\Omega}(x_j', y_j') = \frac{\sum_{i=1}^{n} w_{ij}\, \mu^{NN}(x_i, y_i)}{\sum_{i=1}^{n} w_{ij}}. \tag{6}$$

Thus, source points closer to a target finite element node contribute more strongly to the mapped shear modulus value, whereas distant points have small influence.

We then introduce the IDW mapping into Eq. 8, represented here as $P$, the minimization problem can be rewritten as:

$$\hat{\theta} = \underset{\theta}{\operatorname{argmin}} \left[ \frac{1}{2} \parallel \tilde{\mathbf{u}}' - \tilde{\mathbf{u}}(P(f(\theta, z))) \parallel^2 \right], \quad \hat{\mu} = P(f(\hat{\theta}, z)). \tag{7}$$

### Optimization of the EL-DIP framework

After the network-predicted shear modulus field is mapped onto the finite element mesh by the IDW operator, the EL-DIP reconstruction is obtained by minimizing the displacement-misfit loss in Eq. 7. In this formulation, the predicted displacement field depends on the network parameters indirectly through the finite element equilibrium equations. Therefore, the gradient of the loss function with respect to the network parameters cannot be obtained by a purely explicit differentiation of the network output.

To address this issue, we employ a discretize-then-optimize adjoint strategy[35]. In each optimization iteration, the current shear modulus field is first generated by the neural network and mapped to the problem domain. A finite element forward problem is then solved to compute the displacement field, after which an adjoint problem is solved to efficiently evaluate the gradient of the loss function with respect to the network parameters. The resulting gradient is used by the Adam optimizer to update the network parameters[39].

This adjoint-based back-propagation avoids explicitly forming the sensitivity of the displacement field with respect to all network parameters, and is therefore more efficient for the high-dimensional parameterization used in EL-DIP. The detailed derivation of the adjoint gradient, together with the implementation settings such as the optimizer parameters, learning-rate annealing strategy, iteration numbers, and hardware configuration, is provided in Supplementary Text C.

## Results

### Numerical experiments

In the numerical study, all simulated cases were formulated in normalized units to avoids dependence on a specific physical scale and to facilitate comparison across different experimental scenarios. Accordingly, the geometric dimensions, material parameters, and boundary conditions reported in this subsection are dimensionless unless otherwise stated. Two benchmark geometries were considered: a square domain with side length 2 and a circular domain with radius 1. As seen in Fig. 3, two typical cases with different inclusion configurations were designed in these two domains. The shear modulus values of the background and inclusions were assigned as $\mu_b = 1$ and $\mu_i = 10$, respectively. The boundary conditions are illustrated in Fig. 3a. For the square domain, the bottom boundary was fully fixed and a uniform traction of $t = 1$ was applied on the top boundary. For the circular domain, an inward normalized radial displacement of $u_r = 0.1$ was prescribed along the outer boundary.

The performance of the EL-DIP framework was evaluated against the reconstruction results obtained using the conventional optimization method. For quantitative evaluation, the correlation coedfficient (CC), mean square error (MSE), peak signal-to-noise ratio (PSNR), and stuctural similarity index (SSIM) between the reconstructed shear modulus field $\hat{\mu}$ and the ground-truth field $\mu$ were calculated[40].

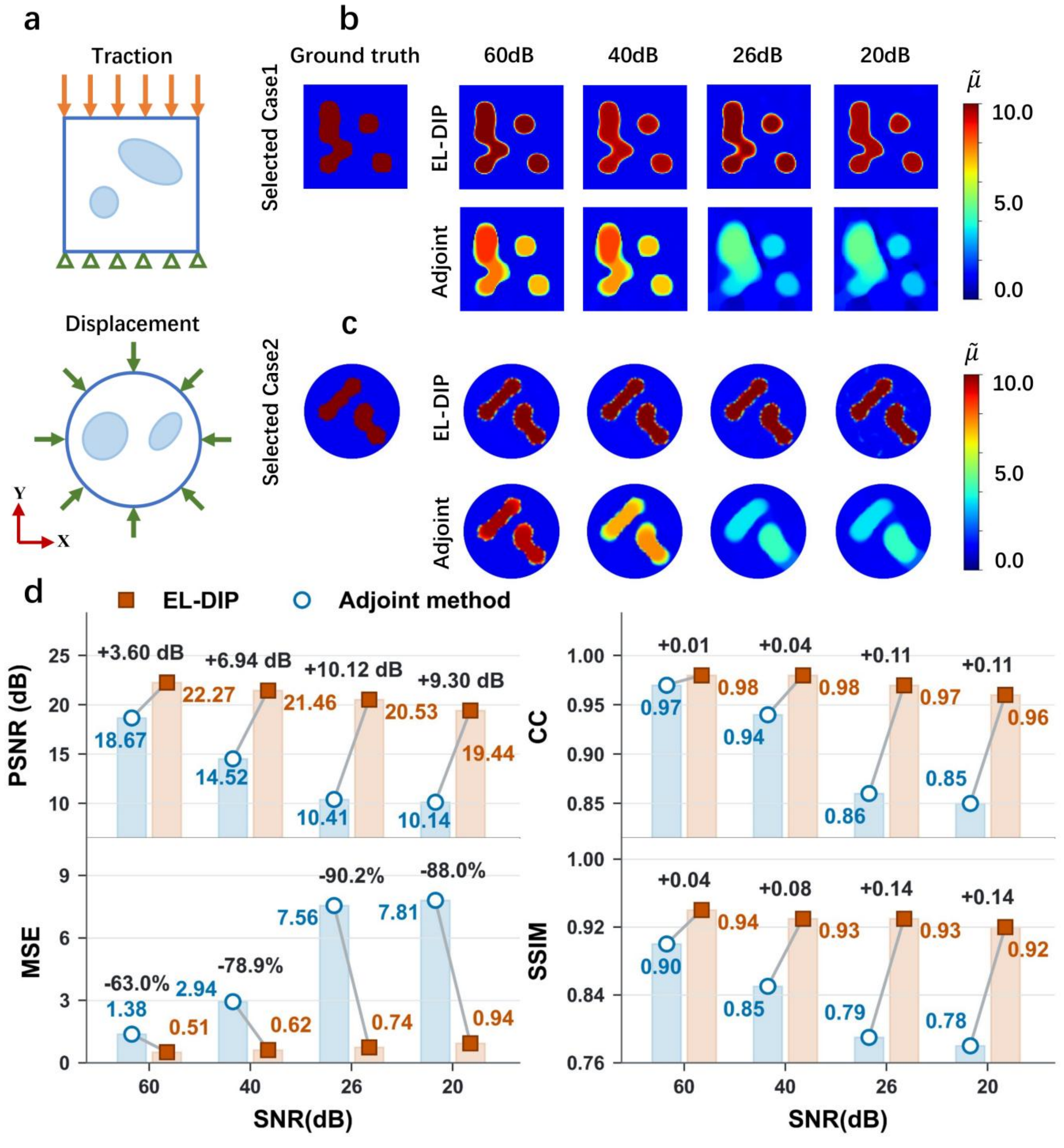

**Fig. 3. Numerical validation and noise robustness of EL-DIP. (a)** Benchmark geometries and loading conditions for the square and circular domains. **(b,c)** Representative reconstructions by EL-DIP and the adjoint optimization (Adjoint) method at SNRs of 60, 40, 26 and 20 dB for the square and circular domains, respectively **(d)** Mean CC, PSNR, MSE, and SSIM across the eight cases. Orange and blue bars denote EL-DIP and Adjoint method, respectively. Black text indicates relative metric gains or MSE reductions. All benchmark quantities are dimensionless.

Here, CC measures the linear correlation between the reconstructed and ground-truth fields, reflecting whether the high- and low-modulus regions are consistently distributed over the whole domain. MSE quantifies the pointwise squared error and therefore directly evaluates the quantitative accuracy of the reconstructed shear modulus values. PSNR further measures the global reconstruction fidelity relative to the dynamic range of the ground-truth shear modulus field and can be regarded as a normalized representation of the reconstruction error. SSIM evaluates the similarity of local mean, contrast, and structural information, and is particularly useful for assessing whether the inclusion geometry, material interfaces, and local structural features are preserved. Therefore, in this study, SSIM is considered especially important for evaluating geometric and interfacial fidelity, whereas MSE and PSNR characterize quantitative modulus accuracy and global fidelity, and CC serves as an auxiliary indicator of overall spatial-distribution consistency. Higher values of CC, PSNR, SSIM and lower MSE, indicate better reconstruction quality. The mathematical definitions of these metrics are provided in Supplementary Text A.

The rectangular and circular domains were discretized using quadrilateral and triangular elements, respectively, with fine meshes comprising 9,216 nodes and 9,025 elements and 2,175 nodes and 4,198 elements, respectively. The reconstructed shear-modulus fields were mapped onto the corresponding physical meshes using inverse distance weighting (IDW) interpolation for visualization and quantitative comparison. To benchmark the performance under more practical conditions, Gaussian noise with SNRs of 60 dB, 40 dB, 26 dB, and 20 dB was introduced to the simulated displacement data.

The detailed case-wise quantitative metrics are provided in Supplementary Text B, while the mean CC, PSNR, MSE, and SSIM values across the eight simulated cases are summarized in Fig. 3d. As the SNR decreases from 60 to 20 dB, both methods exhibit progressive degradation in reconstruction performance. This degradation is substantially more pronounced for the conventional optimization method: CC, PSNR, and SSIM decrease by 12.4%, 45.7%, and 13.3%, respectively. By comparison, EL-DIP exhibits considerably smaller reductions of 2.0%, 12.7%, and 2.1%, respectively, indicating a weaker dependence on measurement noise.

The performance advantage of EL-DIP becomes particularly pronounced under moderate-to-high noise levels. At 26 dB, EL-DIP increases CC from 0.86 to 0.97 and SSIM from 0.79 to 0.93, improves PSNR by 10.12 dB, and reduces MSE by 90.2% relative to the conventional optimization method. Even at 20 dB, EL-DIP retains substantial improvements, with a 9.30-dB gain in PSNR and an 88.0% reduction in MSE. These results indicate that EL-DIP maintains higher reconstruction fidelity as measurement noise increases.

The representative reconstructions in Fig. 3 (b and c), together with the complete comparisons in Figs. S1 and S2, further support the quantitative observations. In both the square and circular domains, EL-DIP preserves the inclusion locations, geometric shapes, and modulus contrast more clearly than the conventional optimization method as the noise level increases. For simple inclusion configurations, both methods can identify the main inclusion regions under low-noise conditions, but the conventional optimization method produces increasingly blurred boundaries and reduced modulus contrast under stronger noise. For more complex inclusion configurations, such as Cases 3,4,6,7, and 8 in Figs. S1 and S2, EL-DIP retains sharper interfaces and more recognizable inclusion geometries, whereas the conventional optimization method suffers from stronger boundary smoothing and loss of structural details. These results suggest that the implicit regularization induced by the DIP architecture is effective in suppressing noise-induced artifacts while preserving the main geometries features of heterogeneous modulus distributions.

## Experimental results

Moreover, to validate the feasibility of the method, we compare EL-DIP with OP in real experimental cases. Subsequently, we acquired experimental data from tissue phantoms in vivo using the Philips iU22 ultrasound system and the L9-3 linear array transducer[41]. The data acquisition process is described in references[42], and a schematic overview is illustrated in Fig. 4a. In the experiment, a soft tissue-mimicking phantom was constructed containing both the inclusions and background regions. The shear moduli of these two components were assigned values of 26.67 kPa and 8.33 kPa, respectively, yielding a shear modulus ratio of 3. The distribution of the inclusions is reconstructed based on the displacement of subregions within the tissue-mimicking phantom.

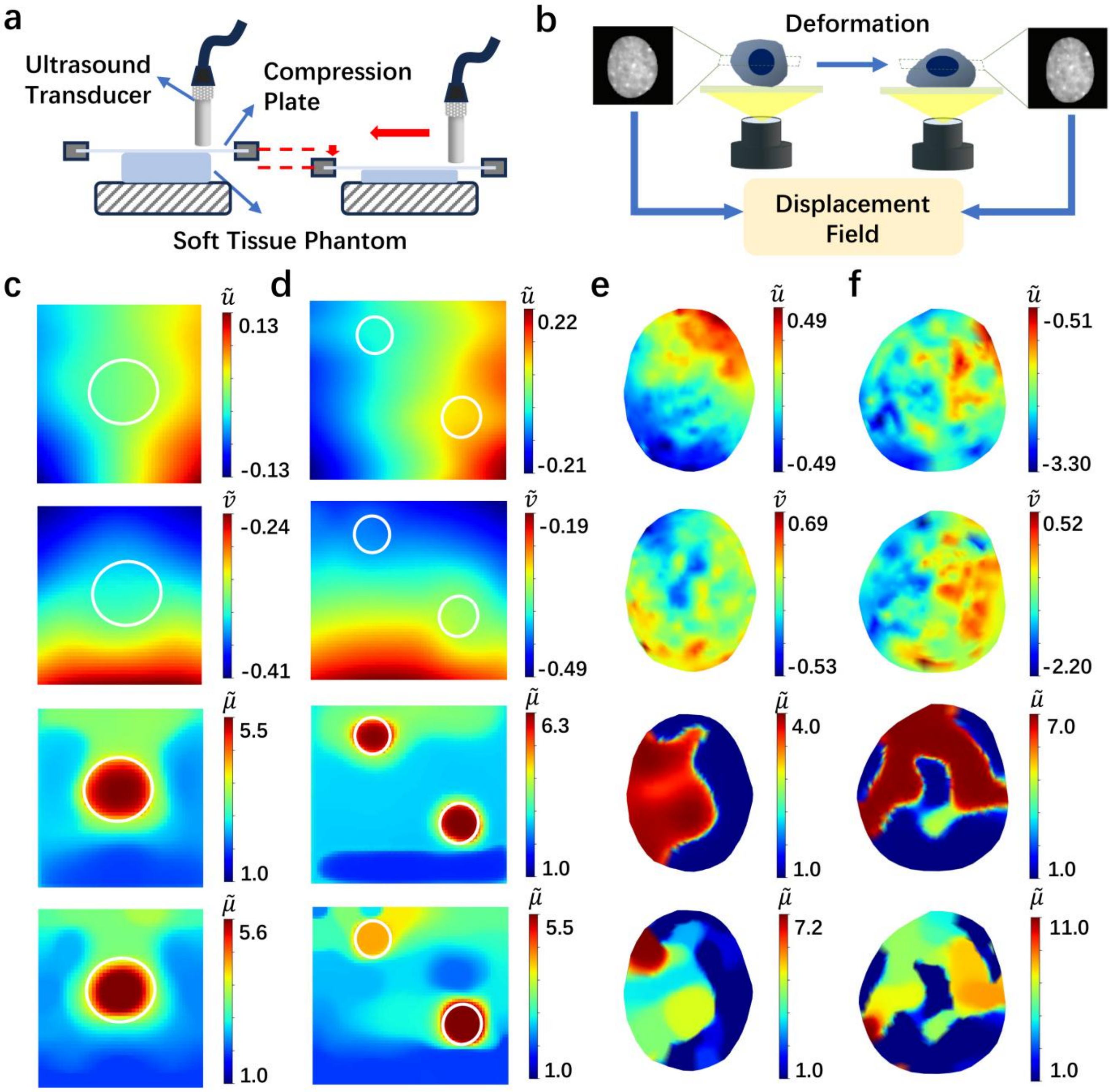

**Fig. 4. Experimental elastography in tissue phantoms and cell nuclei. (a)** Ultrasound-based displacement acquisition using image correlation[43,44]. **(b)** Deformation-microscopy workflow for nuclear displacement measurements[45]. **(c, d)** Single- and double-inclusion tissue phantoms. **(e, f)** Two cardiomyocyte nuclei. In each result column (c to f), rows show, from top to bottom, the horizontal displacement $u$, vertical displacement $v$, EL-DIP shear-modulus reconstruction, and OP reconstruction. Dashed outlines identify phantom inclusions. Individual color scales are retained from the source panels. Panels (e and f) are intentionally repeated in Figs. S3 and S4, together with the original microscopy images.

Displacement field data in both the x- and y-directions for a single inclusion within a sub-region of the tissue-mimicking phantom were obtained using ultrasound imaging technology, as shown in the first two rows of Fig. 4c. The shear modulus distributions reconstructed by EL-DIP and the conventional OP method are presented in the third and fourth rows of Fig. 4c, respectively. As shown in the results, both methods are capable of accurately identifying the shape, location, and shear modulus value of the simple inclusion. However, the EL-DIP method appears to perform better, as the mapped inclusion is closer to a circular shape.

For the two inclusions case (see Fig. 4d), a significantly higher boundary identification accuracy is achieved by the EL-DIP method in complex scenarios characterized by the presence of multiple inclusions. Concurrently, a smoother, more homogeneous background is achieved by the EL-DIP method in the reconstructed images through the suppression of associated artifacts.

Additionally, elastography applied to non-uniform geometries is also explored. The spatial variation of euchromatin and heterochromatin in the nucleus is widely regarded as a critical determinant of gene expression and cellular homeostasis. Consequently, mapping the intranuclear chromatin distribution is deemed highly imperative. In this case, cardiomyocytes were placed on PDMS substrates, and intranuclear planar images were obtained using epifluorescence microscopy. The nuclear displacement field data was derived from the undeformed and deformed images of the nucleus by employing deformation microscopy. The experimental workflow is outlined in Fig. 4b.

We employed the acquired nuclear displacement field data (the first two rows of Fig. 4, e and f) to inversely identify the internal mechanical properties of the nucleus using the EL-DIP and OP methods, respectively. The results are shown in the third and fourth rows of Fig. 4 (e and f), respectively. The original microscopy panels and individual reconstructions are retained in  Figs. S3 and S4. The robustness of the EL-DIP method is demonstrated when confronted with the complex distribution of the two chromatin types within the nucleus, enabling accurate identification of the intranuclear chromatin architecture and clear boundary delineation between euchromatin and heterochromatin. In contrast, the reconstructed results obtained with the OP method do not allow for clear differentiation between the two chromatin types.

## Discussion

The noise robustness of EL-DIP can be partly attributed to the spectral bias of neural networks, namely the tendency of neural networks to learn low-frequency components faster than high-frequency components[46]. This behavior provides an implicit regularization effect: coherent large-scale structures are recovered earlier during optimization, whereas high-frequency noise and oscillatory artifacts are fitted more slowly[30]. In this section, we use neural tangent kernel (NTK) theory to provide an intuitive explanation of this phenomenon[47].

Let the neural network output on a set of discrete input points be denoted by $\mathbf{f}(\theta, \mathbf{x}) = [f(\theta, x_1), f(\theta, x_2), \ldots, f(\theta, x_n)]^T$, and let the target field be $\mathbf{y} = [y_1, y_2, \ldots, y_n]^T$. The loss function is defined as

$$L(\theta) = \frac{1}{2} \parallel \mathbf{f}(\theta, \mathbf{x}) - \mathbf{y} \parallel^2. \tag{8}$$

The gradient-flow dynamics of the network parameters are given by $\frac{\mathrm{d}\theta}{\mathrm{d}t} = -\eta \nabla_\theta L(\theta)$, where $\eta$ is the learning rate. Denote the Jacobian of the network output with respect to the network parameters as $J_\theta = \frac{\partial \mathbf{f}(\theta, \mathbf{x})}{\partial \theta}$. Using the chain rule and substituting the gradient-flow equation gives

$$\frac{\mathrm{d}\mathbf{f}}{\mathrm{d}t} = J_\theta \frac{\mathrm{d}\theta}{\mathrm{d}t} = -\eta J_\theta J_\theta^T (\mathbf{f}(\theta) - \mathbf{y}). \tag{9}$$

The matrix $\mathbf{H}(\theta) = J_\theta J_\theta^T$ is the neural tangent kernel matrix. Its entries are $\mathbf{H}_{ij}(\theta) = \langle \nabla_\theta f(\theta, x_i), \nabla_\theta f(\theta, x_j) \rangle$, where $\langle \cdot,\cdot \rangle$ denotes the inner product in the parameter space. This quantity measures the coupling between the output at $x_i$ and $x_j$ induced by changes in the network parameters.

In the finite-width or small-learning-rate regime, the NTK matrix can be approximately treated as constant during training. The training dynamics then reduce to the following linear system $\frac{\mathrm{d}\mathbf{f}}{\mathrm{d}t} = -\eta \mathbf{H}(\mathbf{f} - \mathbf{y})$.

Let the prediction error be $\mathbf{e}(t) = \mathbf{f}(t) - \mathbf{y}$. Then, the error evolution follows the ode $\frac{\mathrm{d}\mathbf{e}}{\mathrm{d}t} = -\eta \mathbf{H}\mathbf{e}$, whose solution is $\mathbf{e}(t) = \exp(-\eta \mathbf{H} t)\mathbf{e}(0)$.

Since $\mathbf{H}$ is positive semi-definite, it admits the eigendecomposition $\mathbf{H} = \mathbf{Q}\mathbf{\Lambda}\mathbf{Q}^T$, where $\mathbf{Q}$ is an orthogonal matrix composed of eigenvectors, and $\mathbf{\Lambda} = \mathrm{diag}(\lambda_1, \lambda_2, \ldots, \lambda_n)$ contains the corresponding eigenvalues. Projecting the error onto the eigenbasis of $\mathbf{H}$ yields

$$\mathbf{Q}^T \mathbf{e}(t) = \exp(-\eta \mathbf{\Lambda} t)\mathbf{Q}^T \mathbf{e}(0). \tag{10}$$

Therefore, the error component associated with the k-th eigenmode decays as $\exp(-\eta \lambda_k t)$.

This expression shows that eigenmodes associated with larger eigenvalues converge faster, whereas those associated with smaller eigenvalues converge more slowly. For neural networks exhibiting spectral bias, the fast-converging modes are typically dominated by smoother, low-frequency components, while high-frequency components and noise correspond to slower-converging modes. As a result, the network tends to reconstruct coherent large-scale structures before fitting high-frequency noise.

In the EL-DIP framework, this behavior implies that the network-generated shear modulus field tends to capture the dominant inclusion geometry and modulus contrast before fitting noise-induced high-frequency artifacts in the measured displacement data. This provides an implicit regularization mechanism and helps explain why EL-DIP achieves more stable reconstructions under high-noise conditions.

## Conclusion

In this paper, we proposed EL-DIP, a deep-image-prior-based framework for reconstructing heterogeneous material properties from measured deformation fields—a core inverse problem in mechanics. Unlike conventional optimization methods that rely on explicit regularization, EL-DIP parameterizes the unknown property field using an untrained convolutional neural network and exploits the implicit regularization of the network architecture. By integrating IDW-based geometric mapping, an adaptive sigmoid activation layer, finite element forward analysis, and adjoint-based gradient computation, the framework handles arbitrary-shaped domains and efficiently optimizes network parameters without requiring training data.

The effectiveness of EL-DIP was demonstrated through numerical benchmarks and experimental data from soft material phantoms and cell nuclei. The results show that EL-DIP accurately recovers inclusion location, geometry, and modulus contrast, while exhibiting stronger noise robustness than the conventional adjoint method. Under moderate-to-high noise levels, it better preserves inclusion boundaries and suppresses noise-induced artifacts, indicating broad potential for biomechanical characterization, biomedical diagnosis, and non-destructive evaluation. Several limitations remain.

The present study focuses on two-dimensional linear elastic problems, whereas real materials may exhibit nonlinear behavior and three-dimensional deformation. Extending EL-DIP to nonlinear constitutive models and three-dimensional inverse problems is therefore an important direction. In addition, the repeated finite element solutions required during optimization still impose a considerable computational cost. Future work will explore transfer learning and surrogate-model-based acceleration to improve efficiency and practical applicability.

# SUPPLEMENTARY MATERIALS

## EL-DIP: A Deep Image Prior Framework for Accurate, Efficient, and Noise-Robust Elastography

**Hao Liu**[1,2,†], **Fuyi Li**[3, †], **Dong Liu**[4], **Yue Mei**[1,2,*], **Sheng Mao**[3,*]

[1]Department of Engineering Mechanics, Dalian University of Technology, Dalian, 116023, PR China

[2]International Research Center for Computational Mechanics, Dalian University of Technology, Dalian 116023, PR China

[3]School of Mechanics and Engineering Science, Peking University, Beijing 100871, PR China

[4]School of Information Science and Technology, University of Science and Technology of China, Hefei 230093, PR China

†These authors contributed equally to this work

*Corresponding authors: meiyue@dlut.edu.cn, maosheng@pku.edu.cn

## Supplementary Materials

**This PDF file includes:**

## Supplementary Text A. Definitions of quantitative evaluation metrics

In this work, the quantitative evaluation metrics were computed between the reconstructed shear modulus field $\hat{\mu}$ and the ground-truth field $\mu$. Their mathematical definitions are listed below:

$$\mathrm{MSE} = \frac{1}{N}\sum_{i=1}^{N}(\hat{\mu}_i - \mu_i)^2. \tag{S1}$$

$$\mathrm{PSNR} = 10\log_{10}(\frac{\mu_{\max}^2}{\mathrm{MSE}}). \tag{S2}$$

$$\mathrm{CC} = \frac{\sum_{i=1}^{N}(\hat{\mu}_i - \bar{\hat{\mu}})(\mu_i - \bar{\mu})}{\sqrt{\sum_{i=1}^{N}(\hat{\mu}_i - \bar{\hat{\mu}})^2}\sqrt{\sum_{i=1}^{N}(\mu_i - \bar{\mu})^2}}. \tag{S3}$$

$$\mathrm{SSIM}(\hat{\mu}, \mu) = \frac{(2m_{\hat{\mu}}m_{\mu} + C_1)(2\sigma_{\hat{\mu}\mu} + C_2)}{(m_{\hat{\mu}}^2 + m_{\mu}^2 + C_1)(\sigma_{\hat{\mu}}^2 + \sigma_{\mu}^2 + C_2)}. \tag{S4}$$

Here, N denotes the number of pixels or nodes used for the comparison. $\hat{\mu}_i$ and $\mu_i$ are the reconstructed and ground-truth shear modulus values at the i-th location, respectively. $\bar{\hat{\mu}}$ and $\bar{\mu}$ denote the mean values of the reconstructed and ground-truth fields over the whole evaluation domain. $\mu_{\max}$ is the maximum value f the ground-truth shear modulus field and is used as the reference peak value is the PSNR calculation.

For SSIM, $m_{\hat{\mu}}$ and $m_{\mu}$ denote the local mean value of the reconstructed and the ground-truth fields within a sliding window, respectively. $\sigma_{\hat{\mu}}^2$ and $\sigma_{\mu}^2$ denote the corresponding location variances, and $\sigma_{\hat{\mu}\mu}$ denotes the local covariance between the reconstructed and the ground-truth fields. $C_1$ and $C_2$ are small positive constants introduced to avoid numerical instability when the denominators are close to zero. In practice, SSIM was computated over local siding windows, and the final reported SSIM values was obtained by averaging all local SSIM values over the entire field.

Lower MSE and higher PSNR, CC, and SSIM indicate better agreement bwtween the reconstructed shear modulus field and the ground-truth field.

## Supplementary Text B. Case-wise quantitative results of simulated examples

Supplementary Text B provides the detailed case-wise quantitative results corresponding to the numerical examples discussed in the Simulated results subsection. Table S1 lists the values of CC, PSNR, MSE, and SSIM for all eight simulated cases under four different noise levels, namely 60 dB, 40 dB, 26 dB, and 20 dB. These metrics were computed by comparing the reconstructed shear modulus field with the corresponding ground-truth field on the same evaluation discretization. Higher values of CC, PSNR, and SSIM indicate better reconstruction performance, whereas a lower MSE is preferred. For each case, noise level, and metric, the better value between EL-DIP and the conventional optimization method is highlighted in bold. The mean values over the eight cases are summarized in the main text to provide an overall comparison between the two methods.

## Supplementary Text C. Adjoint gradient derivation and implement details

### Back-propagation based on the adjoint method

The key step in the optimization is to evaluate the gradient of the loss function $J$ with respect to the network parameters $\theta$ under the constraint of the governing equilibrium equations in the main text. Following a discretize-then-optimize strategy, we first discretize the mechanical problem and then apply the adjoint method to the resulting algebraic system. Let $\mathbf{u}$ denote the vector of nodal displacement unknowns, and let the discretized equilibrium equation be written as:

$$\mathbf{G}(\mathbf{u}, \theta) = 0. \tag{S5}$$

Within the discrete finite element function space, the gradient of loss function J can be determined by the following relation:

$$\frac{\mathrm{d}J}{\mathrm{d}\theta} = \frac{\partial J}{\partial \mathbf{u}}\frac{\mathrm{d}\mathbf{u}}{\mathrm{d}\theta} + \frac{\partial J}{\partial \theta}. \tag{S6}$$

where the calculation of $\frac{\partial J}{\partial \mathbf{u}}$ and $\frac{\partial J}{\partial \theta}$ is straightforward. On the other hand, the parameters $\theta$ and the displacement field $\mathbf{u}$ are implicitly related via $\mathbf{G}(\mathbf{u}, \theta) = 0$. By differentiating both sides with respect to $\theta$, we have the following relation:

$$\frac{\partial \mathbf{G}}{\partial \mathbf{u}}\frac{\mathrm{d}\mathbf{u}}{\mathrm{d}\theta} = -\frac{\partial \mathbf{G}}{\partial \theta}. \tag{S7}$$

therefore, by chain rule, we have:

$$\frac{\mathrm{d}J}{\mathrm{d}\theta} = -\underbrace{\frac{\partial J}{\partial \mathbf{u}}\left(\frac{\partial \mathbf{G}}{\partial \mathbf{u}}\right)^{-1}}_{\lambda^*}\frac{\partial \mathbf{G}}{\partial \theta} + \frac{\partial J}{\partial \theta}. \tag{S8}$$

where $\lambda$ is from solving adjoint equation:

$$\left(\frac{\partial \mathbf{G}}{\partial \mathbf{u}}\right)^{*}\lambda = \left(\frac{\partial J}{\partial \mathbf{u}}\right)^{*}. \tag{S9}$$

Such approach is often called the adjoint method. Since the dimension of the parameters $\theta$ is much larger than the scalar loss function $J$, the adjoint method is more efficient than directly solving Eq. S7. The computed gradient is then used to update the network parameters $\theta$ using the Adam optimizer.

## Implementation details

The flowchart of the EL-DIP is illustrated in the main text. We use the Adam optimizer to optimize the network parameters, with its learning rate set to $lr = 0.1$, momentum parameters $\beta_1 = 0.5$, and $\beta_2 = 0.999$. Here, the maximum number of iterations was set to 2000 for the simulated data and 500 for the experimental data, respectively. During the EL-DIP training process, divergence sometimes occurs. To resolve the issue, we employ the method of learning rate annealing. We will regularly track the average Loss and current Loss of the network. If the value of $(Loss - Loss_{avg})/Loss$ is greater than 0.1, we will restore the parameters to the corresponding parameters of the previous iteration and anneal the learning rate by a factor of 0.9. In this study, all training and testing were performed using PyTorch 2.8.0, an Intel i9-13950HX processor, and an NVIDIA RTX 5000 Ada GPU.

**Table S1. Case-wise quantitative reconstruction metrics under different noise levels.** CC, PSNR, MSE, and SSIM are computed between the reconstructed shear modulus field and the ground-truth field. Bold values indicate the better performance between EL-DIP and the conventional optimization method for each case, noise level, and metric.

| | | 60dB | | | | 40dB | | | | 26dB | | | | 20dB | | | |
|---|---|---|---|---|---|---|---|---|---|---|---|---|---|---|---|---|---|
| | | CC | PSNR | MSE | SSIM | CC | PSNR | MSE | SSIM | CC | PSNR | MSE | SSIM | CC | PSNR | MSE | SSIM |
| **Case1** | EL-DIP | **0.99** | **25.01** | **0.26** | **0.95** | **0.99** | **24.25** | **0.30** | **0.95** | **0.98** | **21.36** | **0.59** | **0.94** | **0.97** | **19.36** | **0.94** | **0.94** |
| | OP | 0.97 | 17.74 | 1.36 | 0.87 | 0.96 | 16.06 | 2.01 | 0.84 | 0.92 | 11.26 | 6.07 | 0.78 | 0.91 | 11.11 | 6.28 | 0.75 |
| **Case2** | EL-DIP | **0.99** | **24.69** | **0.28** | **0.95** | **0.98** | **23.70** | **0.35** | **0.94** | **0.98** | **22.57** | **0.45** | **0.94** | **0.97** | **20.95** | **0.65** | **0.93** |
| | OP | 0.94 | 14.29 | 3.02 | 0.85 | 0.94 | 14.23 | 3.06 | 0.85 | 0.92 | 12.74 | 4.31 | 0.82 | 0.89 | 11.20 | 5.17 | 0.77 |
| **Case3** | EL-DIP | **0.98** | **22.94** | **0.41** | **0.92** | **0.98** | **21.44** | **0.58** | **0.91** | **0.97** | **20.44** | **0.73** | **0.90** | **0.96** | **19.27** | **0.96** | **0.90** |
| | OP | 0.96 | 15.35 | 2.36 | 0.81 | 0.95 | 14.02 | 3.21 | 0.76 | 0.92 | 10.47 | 7.26 | 0.73 | 0.91 | 10.24 | 7.66 | 0.71 |
| **Case4** | EL-DIP | **0.98** | **21.42** | **0.58** | **0.91** | **0.97** | **20.34** | **0.75** | **0.89** | **0.97** | **20.70** | **0.69** | **0.89** | **0.96** | **19.27** | **0.96** | **0.89** |
| | OP | 0.95 | 16.44 | 1.84 | 0.84 | 0.95 | 15.77 | 2.15 | 0.82 | 0.80 | 10.18 | 7.77 | 0.69 | 0.80 | 10.11 | 7.90 | 0.68 |
| **Case5** | EL-DIP | 0.98 | 21.06 | 0.63 | 0.94 | **0.98** | **20.94** | **0.65** | **0.94** | **0.97** | **20.05** | **0.80** | **0.94** | **0.96** | **19.17** | **0.98** | **0.92** |
| | OP | **0.99** | **22.92** | **0.41** | **0.96** | 0.94 | 14.81 | 2.64 | 0.89 | 0.84 | 9.71 | 8.67 | 0.81 | 0.84 | 9.66 | 8.77 | 0.81 |
| **Case6** | EL-DIP | 0.97 | **20.56** | **0.71** | **0.95** | **0.97** | **19.35** | **0.94** | **0.93** | **0.96** | **18.81** | **1.06** | **0.93** | **0.95** | **18.12** | **1.25** | **0.94** |
| | OP | **0.98** | 20.18 | 0.78 | **0.95** | 0.91 | 13.53 | 3.60 | 0.89 | 0.81 | 9.85 | 8.39 | 0.82 | 0.81 | 9.86 | 8.36 | 0.82 |
| **Case7** | EL-DIP | 0.97 | 20.56 | 0.71 | **0.95** | **0.97** | **19.96** | **0.82** | **0.94** | **0.97** | **19.58** | **0.89** | **0.94** | **0.96** | **18.95** | **1.03** | **0.93** |
| | OP | **0.98** | **22.04** | **0.51** | **0.95** | 0.95 | 14.89 | 2.63 | 0.89 | 0.85 | 9.57 | 8.95 | 0.81 | 0.85 | 9.44 | 9.22 | 0.80 |
| **Case8** | EL-DIP | **0.99** | **21.92** | **0.52** | 0.94 | **0.98** | **21.66** | **0.55** | **0.95** | **0.98** | **20.71** | **0.69** | **0.94** | **0.97** | **20.46** | **0.73** | **0.92** |
| | OP | 0.98 | 20.42 | 0.74 | **0.95** | 0.93 | 12.87 | 4.18 | 0.88 | 0.80 | 9.51 | 9.07 | 0.83 | 0.81 | 9.51 | 9.08 | 0.83 |

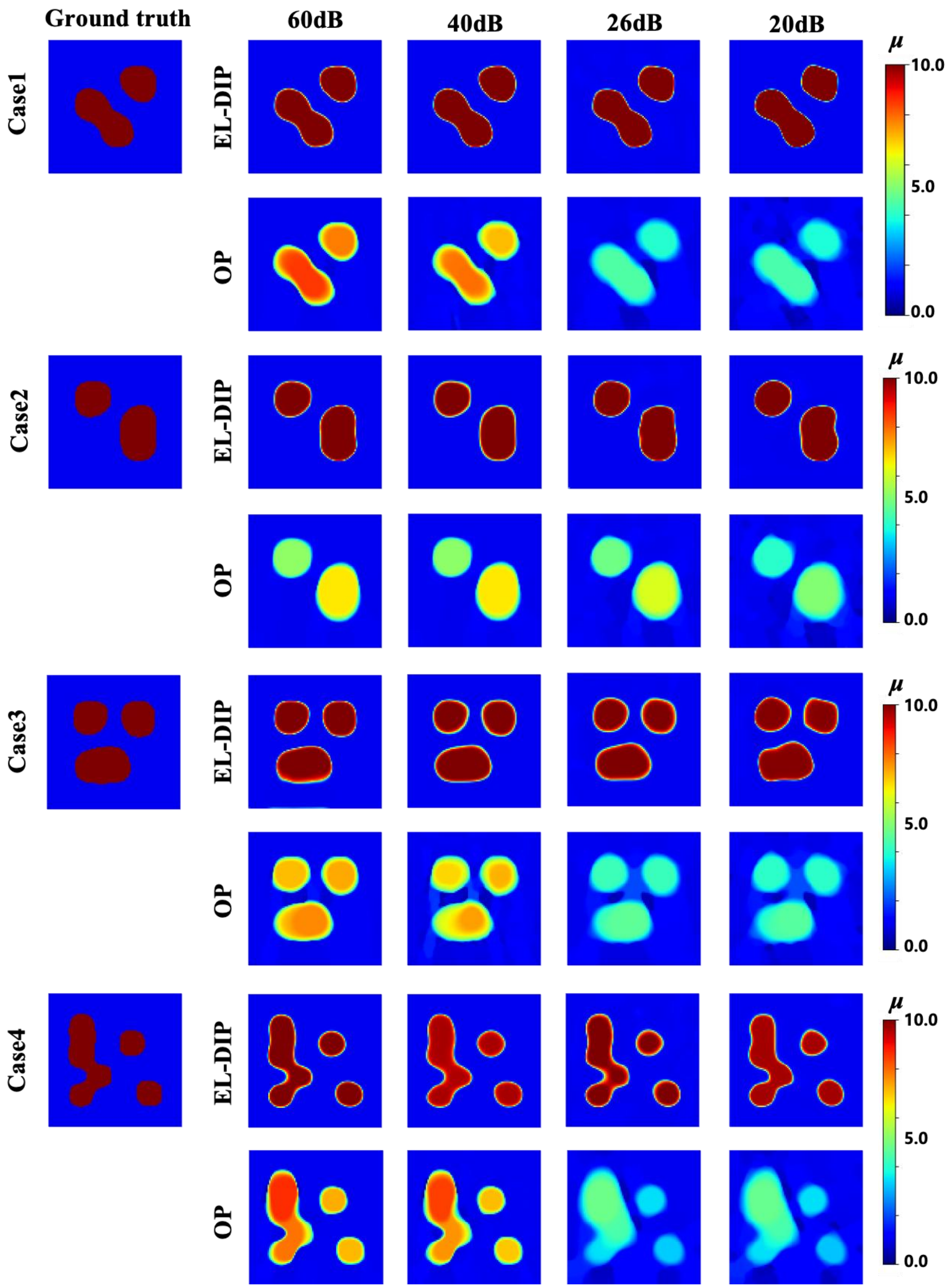

**Fig. S1. Reconstruction results in the rectangular domain under different noise levels.** Comparisons are shown between EL-DIP and the conventional optimization method for representative inclusion configurations and noise conditions. Selected reconstruction panels are intentionally repeated in main-text Fig. 3b.

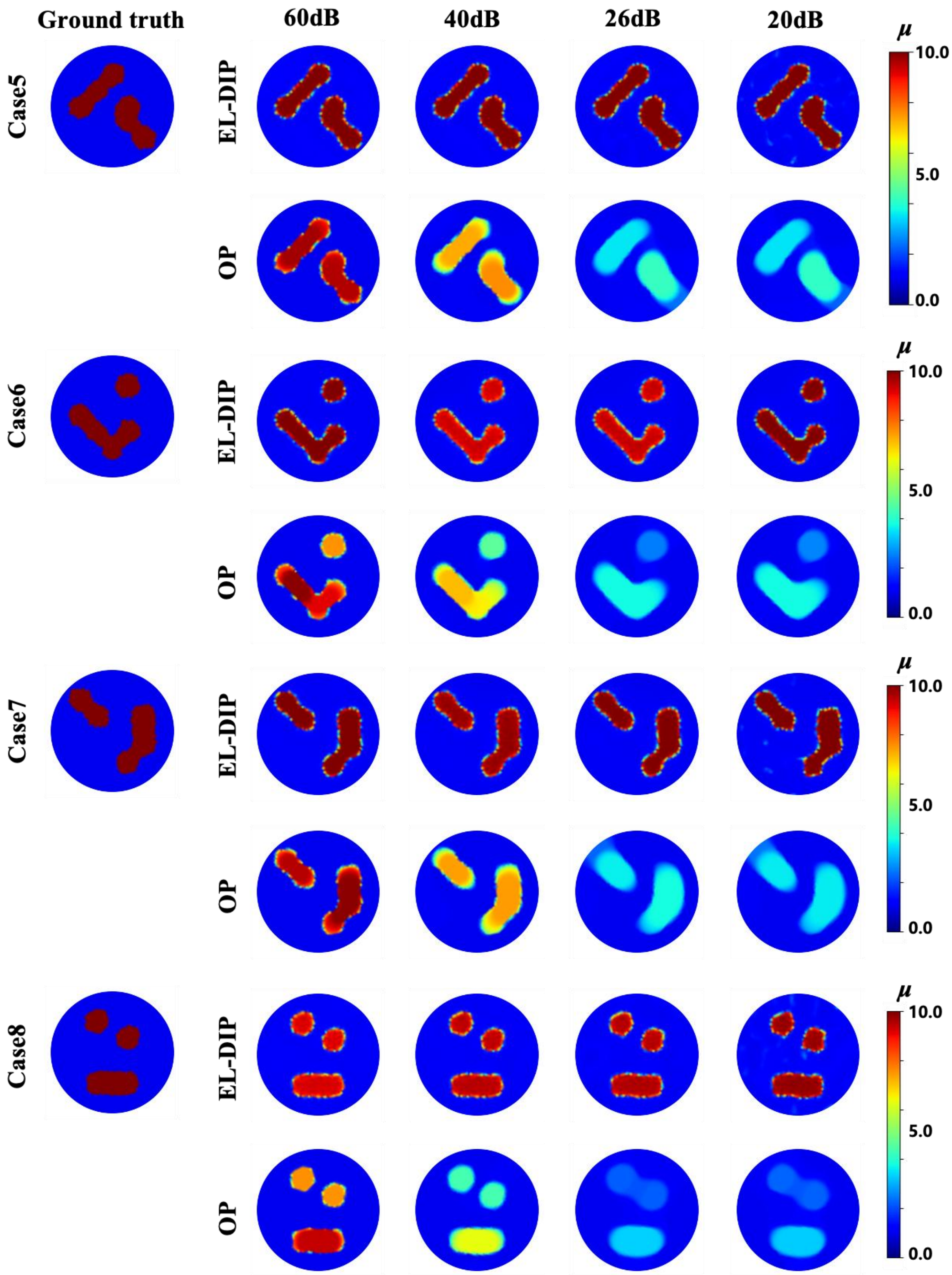

**Fig. S2. Reconstruction results in the circular domain under different noise levels.** Comparisons are shown between EL-DIP and the conventional optimization method for representative inclusion configurations and noise conditions. Selected reconstruction panels are intentionally repeated in main-text Fig. 3c.

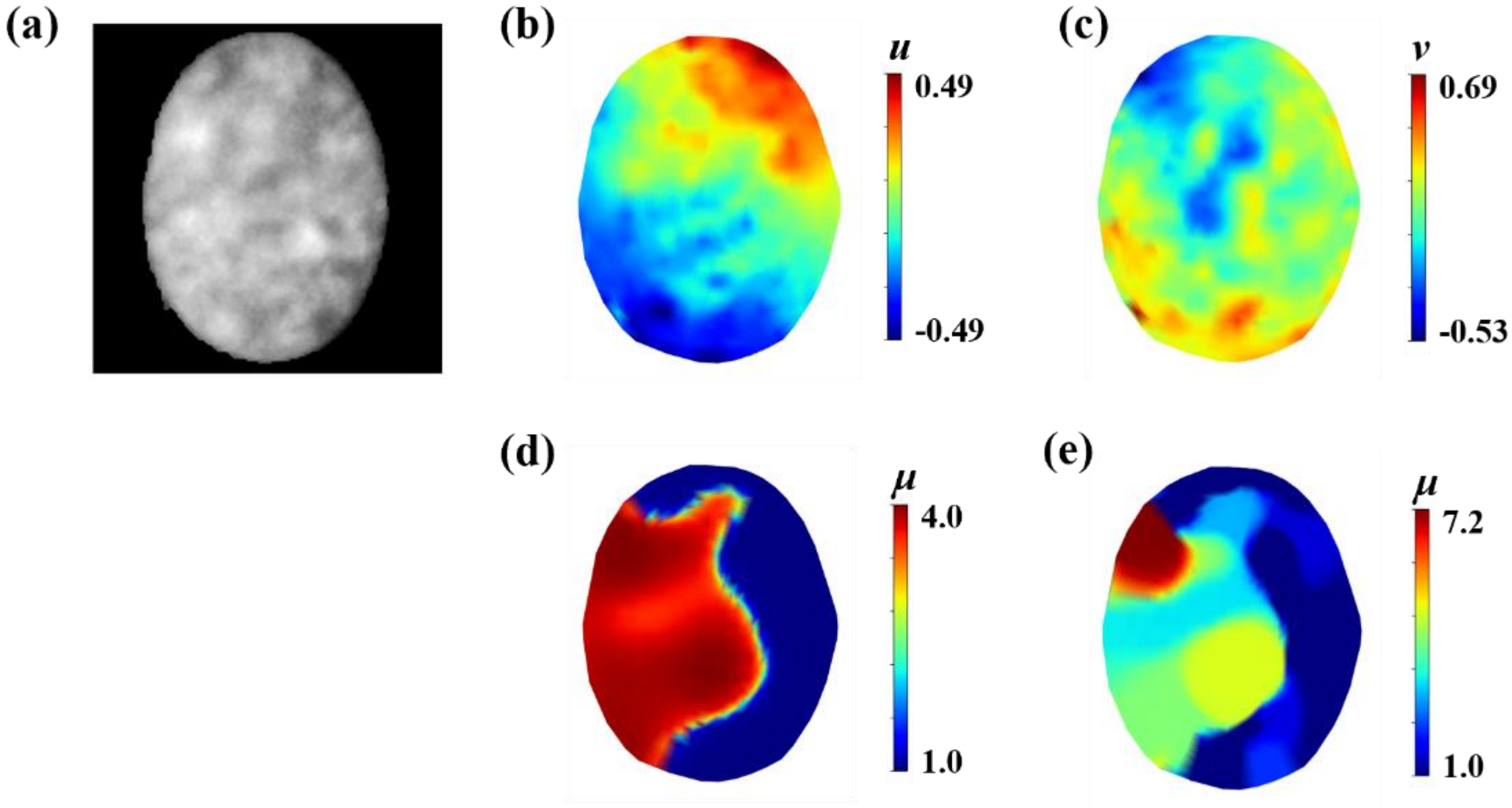

**Fig. S3. Intranuclear elastography result for the first cardiomyocyte nucleus. a,** Microscopic image of the internal plane of the cell nucleus. **b,** Displacement field in the x-direction of the cell nucleus. **c,** Displacement field in the y-direction of the cell nucleus. **d,** Shear-modulus reconstruction obtained by EL-DIP. **e,** Shear-modulus reconstruction obtained by the conventional optimization method. The displacement and reconstruction panels are also included in main-text Fig. 4; the original microscopic image is retained here.

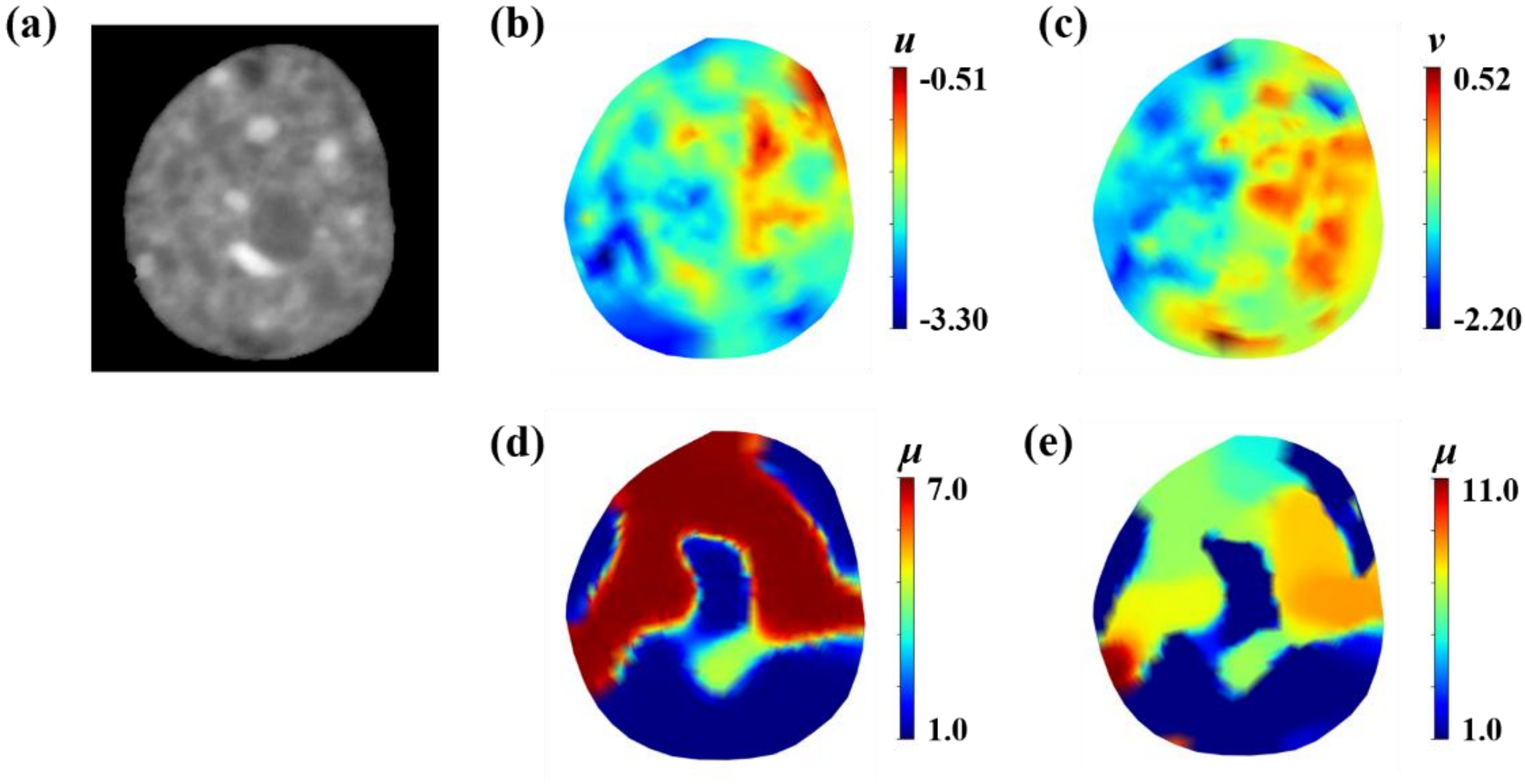

**Fig. S4. Intranuclear elastography result for the second cardiomyocyte nucleus. a,** Microscopic image of the internal plane of the cell nucleus. **b,** Displacement field in the x-direction of the cell nucleus. **c,** Displacement field in the y-direction of the cell nucleus. **d,** Shear-modulus reconstruction obtained by EL-DIP. **e,** Shear-modulus reconstruction obtained by the conventional optimization method. The displacement and reconstruction panels are also included in main-text Fig. 4; the original microscopic image is retained here.